\documentclass[mypaper,7pt,twoside]{CoAst}
\usepackage{epsf,graphicx,fancyhdr}
\usepackage[]{amsmath,amsfonts,amssymb,amsxtra}
\usepackage{ulem}
\renewcommand{\chaptermark}[1]{\markboth{#1}{}}

\newcommand{\logg}{\ensuremath{\log g}}                     

\newcommand{\kopf}{\small\itshape Comm. in Asteroseismology \\ Contribution to the Proceedings of the 38$^{th}$\,LIAC\,/\,HELAS-ESTA\,/\,BAG, 2008
}

\newcommand{\Authors}[1]{\begin{center}\normalsize\bf\sf #1 \end{center}}

\renewcommand{\author}[1]{\begin{center}\normalsize\bf\sf #1 \end{center}}
\newcommand{\Address}[1]{\begin{center}\small\sf #1 \end{center}}

\DeclareGraphicsExtensions{.eps,.jpg}

\newcommand{\Session}[1]{{\vspace{3mm}\small \noindent  \hspace*{3mm} Session: } #1 \normalsize}

	\newcommand{\one}{\small Physics and uncertainties in massive stars on the MS and close to it}

\renewenvironment{abstract}{\section*{Abstract}\normalsize\sf}{}
\newcommand{\References}[1]{\begin{flushleft}{\large References\\}\vspace*{2mm}\small #1 \end{flushleft}}

\newcommand{\chapterCoAst}[2]{\chapter[\sf\normalsize #1\\ \footnotesize \hspace*{5mm}by #2 \sf\normalsize][]{#1\\}\rhead[\fancyplain{}{\sf\footnotesize \center{#1}}]{\fancyplain{}{\sffamily\thepage}}\lhead[\fancyplain{\kopf}{\sffamily\thepage}]{\fancyplain{\kopf}{\sf\footnotesize \center{#2}}}}

\newcommand{\figureCoAst}[5]{\begin{figure}[#4]
\centering
\includegraphics*[#5]{#1}
\caption{#2}
\label{#3}
\end{figure}}

\renewcommand{\chaptername}{}

\newcommand{\acknowledgments}[1]{\vspace*{5mm}\noindent  \textbf{Acknowledgments.} #1}

\def\rfr{\smallskip\par\noindent
        \hangindent=7truemm
        \hangafter=1}

\begin{document}
\sf

\chapterCoAst{ Rotational Mixing in Magellanic Clouds B Stars - Theory versus Observation}
{I.\,Brott, I.\,Hunter, A.\, de Koter, N.\,Langer, D.\,Lennon,  and P.\,Dufton} 

\Authors{I.\,Brott$^{1}$, I.\,Hunter$^{2}$,  A.\, de Koter$^{1,4}$, N.\,Langer$^{1}$, D.\,Lennon$^{3}$ and P.\,Dufton$^{2}$} 

\Address{$^1$ Sterrenkundig Instituut Utrecht, Universiteit Utrecht, Princetonplein 5,
3584CC Utrecht, The Netherlands\\
$^2$ Astrophysics Research Centre, School of Mathematics \& Physics, The
Queen's University of Belfast, Belfast, BT71NN, Northern Ireland,UK \\
$^3$ Space Telescope Science Inst., 3700 San Martin Drive, Baltimore, MD 21218, USA\\
$^4$ Astronomical Institute Anton Pannekoek, University of Amsterdam, Kruislaan 403, 1098SJ Amsterdam, The Netherlands}

\noindent
\begin{abstract}
  We have used VLT FLAMES data to constrain the physics of rotational mixing in stellar evolution models.  We have simulated a
  population of single stars and find  two groups of observed stars that cannot be explained: (1) a group of fast rotating  stars which do not show
  evidence for rotational mixing and (2) a group  of slow rotators with strong N enrichment. Binary effects and fossil magnetic fields may be
  considered to explain those two groups. \\
  We suggest that the element boron could be used to distinguish between rotational mixing and the binary scenario. Our single star population
  simulations quantify the expected amount of boron in fast and slow rotators and allow a comparison with measured nitrogen and boron abundances in B-stars.
\end{abstract}

\Session{ \one } 

\subsection*{Introduction}
Rotational mixing is thought to be an important process in massive stars. It can affect surface abundances and  the internal structure of the
star. The VLT FLAMES Survey (Evans et al. 2005) produced nitrogen abundances for a large sample of B stars, providing valuable empirical
constraints for this  mixing process. 
Nitrogen is an easily observed tracer element for rotational mixing. It is produced in the stellar center and can be mixed gradually to the surface over the main sequence lifetime.  So, in fast rotating stars one expects N enhancements towards the end of the main sequence, while slow rotators should not show any N enhancement. To compare
predictions from our stellar evolution models to the observations from Hunter et al. (2008a) (hereafter FLAMES data), we simulate a population of
single stars to which we apply the same selection criteria. We limit ourselves here to an analysis of main sequence (MS) stars in the Magellanic Clouds.

\figureCoAst{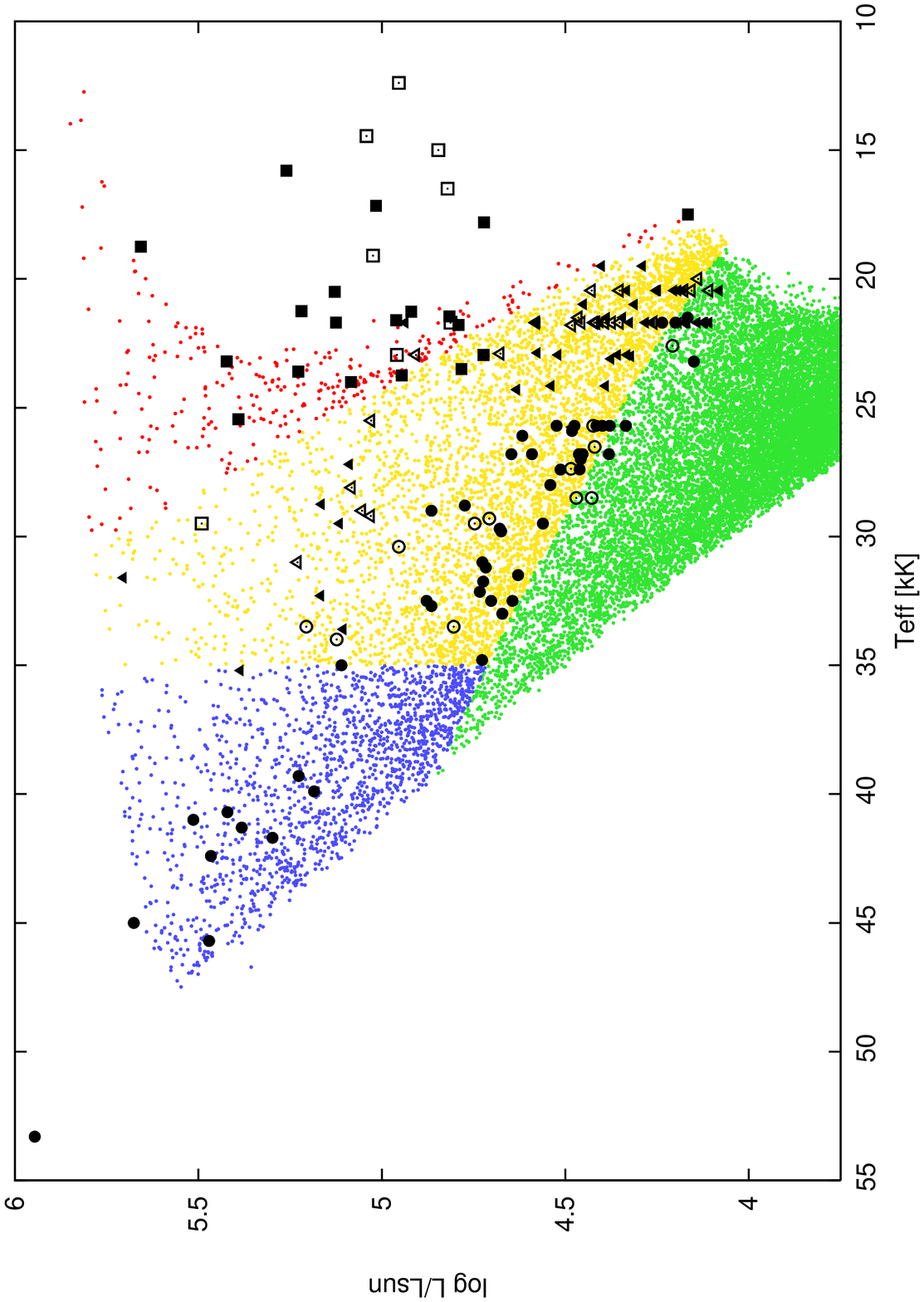}{Result of a population synthesis calculation for $10^6$ stars between 5 and 50 $\rm{M}_\odot$ for LMC composition in
  the HR diagram. Overploted are the observed FLAMES data. Circles are stars with  $\log g > 3.7$\,dex (i.e. early MS stars),
  triangles are stars with $3.2 \leqslant \log g \leqslant 3.7$\,dex (i.e. late MS stars), squares symbolize stars with $\logg <
  3.2$\,dex. Open/filled symbols represent probable binaries/single stars, respectively. The colored regions are explained in the text. Only the
  yellow region is considered in the analysis below.}{selectioneffects}{t!}{clip,angle=-90,width=93mm} 

\subsection*{The Model Grid and Population Simulation}
As input for the population simulation we have  calculated a grid of more than 500 evolutionary sequences until core hydrogen exhaustion. The models
have been calculated with the stellar evolution code BEC (Langer 1991; Heger et al. 2000). The models include rotational mixing,
angular momentum transport by magnetic fields and theoretical mass loss rates of Vink et al. (2001). Rotational mixing and core overshooting have been
calibrated at an evolutionary model representing the average mass and rotation velocity of the LMC sample (Hunter et al. 2008b). The mass and initial
velocity range of the model grid have been chosen to cover most of the FLAMES data, i.e $5-50\,\rm{M}_\odot$, $0-500$\,km/s. It covers three
metallicities, using chemical compositions for SMC, LMC and Galactic evolution models (Brott et al. 2009, in preparation). As input chemical
composition for our stellar models we used C,N,O abundances from H{\tt II} regions (Kurt \& Dufour 1998) and Mg, Si and Fe from unenriched B-stars.
The abundances for remaining elements have been taken from Asplund et al. (2005) and lowered by 0.4 and 0.7 dex for the LMC and SMC, respectively.\\
By comparing isochrones to the LMC and SMC FLAMES data we find a wide age spread, consistent with a constant star formation rate for both observation
sets. This is in agreement with most of the observed B-stars being field stars (Hunter et al. 2008c). We assume a Salpeter initial mass
function. Since the initial velocity distribution of the stars is unknown, we assume that the observed distribution is close to the initial one. Mass
loss is generally small for MS B-stars at LMC metallicity, justifying this approximation.  For each simulated star an age, initial mass, initial
rotational velocity and a random inclination is drawn from the appropriate distribution functions.\\
  Fig. \ref{selectioneffects}  shows an HR diagram of the observed and simulated data points of the LMC. In order that our simulations are consistent
  with the FLAMES observational dataset, we have excluded simulations that fall below the FLAMES magnitude limit (green) or that have $\rm{T_{eff}} > 35$\,kK. The
  analysis of the hotter O-type sample (blue) by Mokiem et al. (2007) has not
  been considerd as no Nitrogen abundance estimates were available (blue). We
  have excluded post-MS stars, which we define as having $\log g < 3.2$\,dex (red) (Hunter et al. 2008b). This gravity limit is appropriate for models
  with initial masses of $13\,\rm{M}_\odot$  but may be too high for higher mass models due to the large overshooting parameter ($\alpha$ = 0.335)
  in the models. We note that the observed points that we have excluded are all slow rotators and hence will not significantly affect our
  conclusions. In the next section we compare all remaining data points (yellow) to the FLAMES data.

\figureCoAst{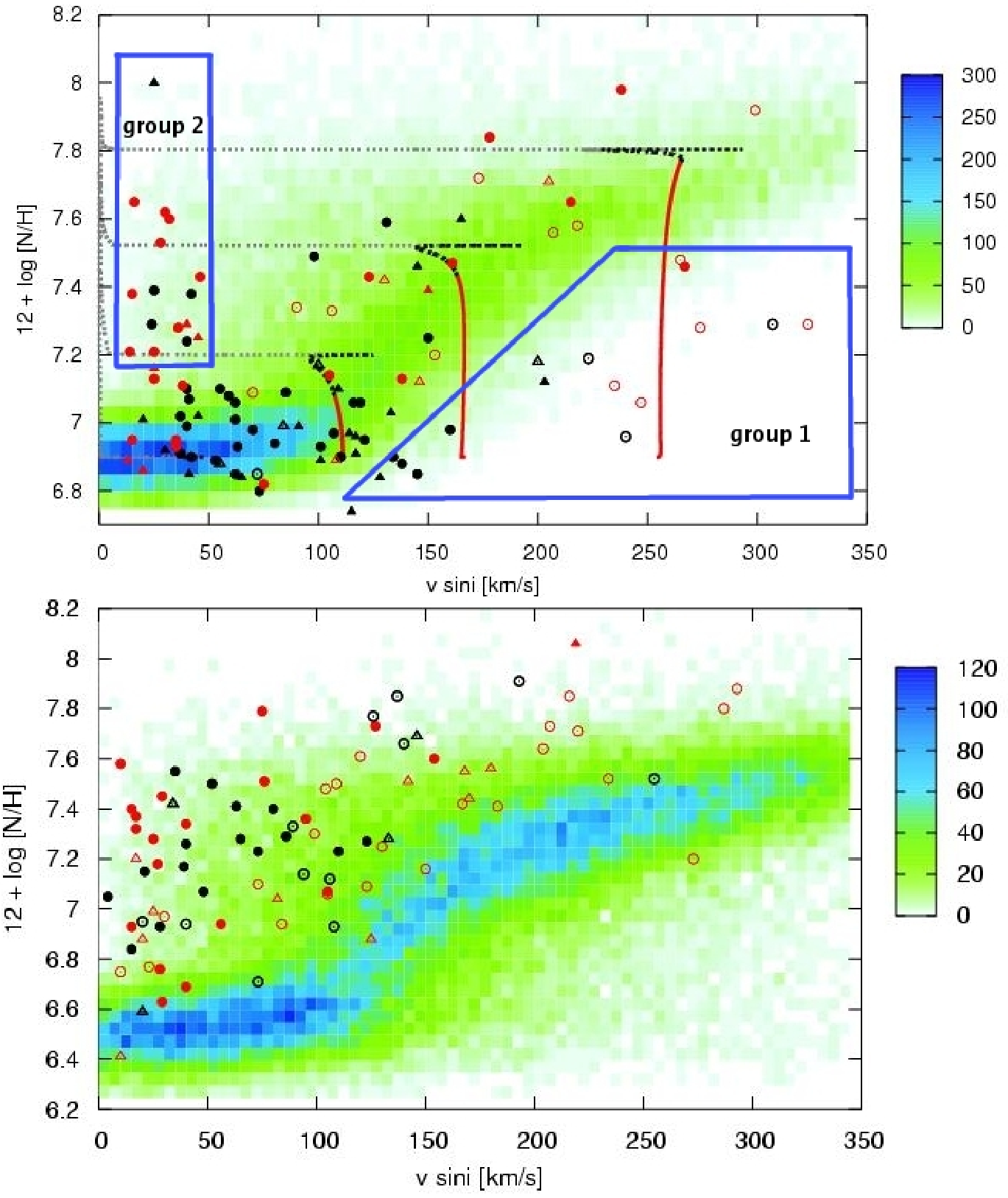}{Surface nitrogen abundance as a function of projected rotational velocity for the LMC (top) and SMC
  (bottom). Observations and a set of 13 $\rm{M}_\odot$ evolution tracks for different  initial rotation velocities are compared. Red (dark gray)
  symbols are lower MS objects ( $\log g > 3.7$), the corresponding parts on the tracks are marked in the same color. Black symbols are upper MS
  objects ($3.2 \leqslant \log g \leqslant 3.7$), the corresponding part on the track is marked in the same color. Triangles show possible  binaries,
  circles are single stars and open symbols are upper limits to the nitrogen abundance. The  shading in the background gives the number of simulated
  stars in this field, as shown by the bars on the right.}{nitrogen}{p}{clip,angle=0,width=110mm} 
 
\subsection*{Nitrogen in the LMC and SMC}
 We have performed a simulation of $10^6$ stars between 5 and 50 $\rm{M}_\odot$ for both Magellanic Clouds. Only data  passing the  observational
 selection criteria (LMC $\sim 4.1\%$, SMC $\sim 4.9\%$) were kept. For better visibility we have binned the data, shown in the density plots in
 Fig. \ref{nitrogen}. The evolutionary tracks in Fig. \ref{nitrogen} top (LMC) show that nitrogen is  mixed gradually to the  surface during the MS
 evolution. The population simulation for single stars shows for both the LMC (top) and SMC (bottom) that, while in fast rotating stars a
 significant enrichment in N is expected, slow rotators show almost no enhancement. While this is true for most of the LMC FLAMES data, there are two
 groups that do not agree with the simulation results. At high $v\sin i $ and low N, a group of mainly upper limits is populating the diagram (group
 1). The simulation can not populate this region, because fast rotating and non-enriched stars  are very young and close to the Zero Age Main Sequence
 (ZAMS). Thereby they are too faint to pass the survey magnitude limit. These stars could have been spun up in a binary system, but only two objects
 in this group are identified radial velocity variables. Group 2 (low $v\sin i$ and highly N enriched) is also in clear contradiction to the
 predictions. 
 These stars can not all be pole on stars, but  are intrinsically slow rotators. Morel et al. (2008) have shown for a Galactic sample of slowly rotating N-rich stars
 that a large fraction of them is magnetic. This suggests that internal magnetic fields might give rise to the abundance anomalies
 of this  group.

 About 50\% of the FLAMES sample are consistent with our  single-star models
 with rotational mixing (see Maeder et. al, this volume), however neither set of models can explain groups 1 and 2. In fact
the binary models of Langer et al. (2008) show promise of accounting for these groups and, to add further ambiguity
to the interpretation, can explain the so-called normal stars as well.
  \\
In the SMC (Fig. \ref{nitrogen}, bottom) almost all observed stars seem to have higher nitrogen enrichments than predicted by the models, even though
the baseline is in agreement with the most unenriched stars in the sample. However, almost all faster rotators ($ v \sin i > 100$\,km/s) have upper
limits, hence are consistent with the predictions. As suggested by Hunter et al. 2008a the enriched group at low velocities might be the analogue of
group 2 in the LMC. In both cases, the group consists mainly of high log g objects.
 
\subsubsection*{Boron to Discriminate between a Binary and Single Star Scenario}
The true binary fraction in the sample is of key importance to understand the groups that are in disagreement with the single star simulation. The
element boron can only exist in the coolest outer layers of the star. It  will be gradually destroyed by rotational mixing as shown in
Fig. \ref{bor_lmc}. In a mass transfer scenario practically boron free material is dumped onto the mass gainer. Thus, in a pure binary sample,  boron
is either almost at its initial abundance or it is significantly depleted by mass transfer (Fliegner et al. 1996). Boron abundances therefore will
help identify effects of single versus binary stars in diagnostics such as Fig.\,\ref{nitrogen}, even though Boron measurements in the Magellanic Clouds are still very challenging at present.

\figureCoAst{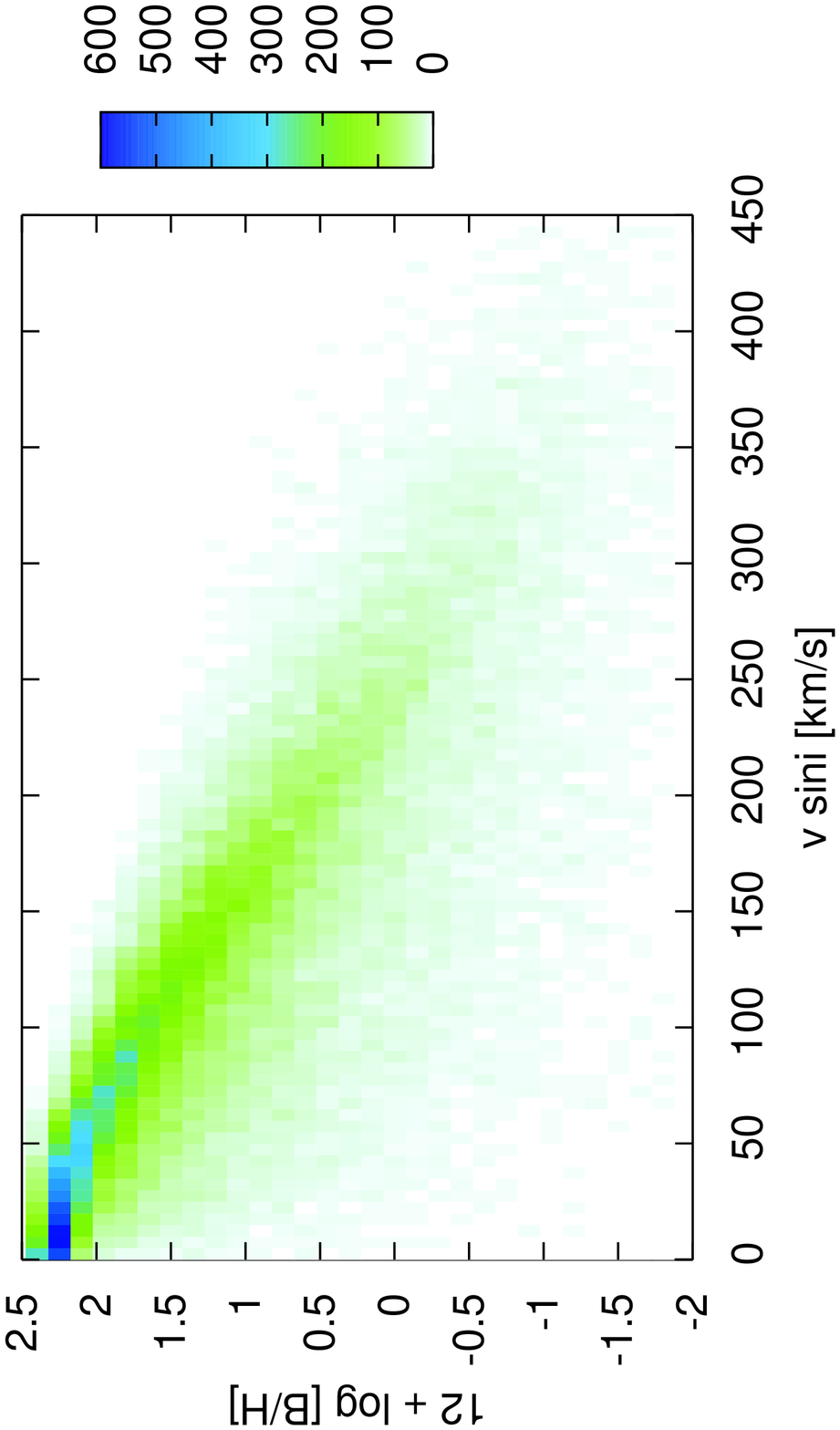}{Predicted boron vs.  rotational velocity for a population of single stars in the LMC. The trend
  reflects the transport of this element to hot deep layers, through rotational mixing, where it is subsequently
  destroyed.}{bor_lmc}{!t}{clip,angle=-90,width=90mm}  

\acknowledgments{Thanks to Peter Anders for many helpful discussions.}

\References{
\rfr Asplund, M., Grevesse, N., \& Sauval, A. 2005, ASPC,336,25
\rfr Evans, C.J., Smartt, S.J., Lee, J-K., et al. 2005, A\&A, 437, 467
\rfr Fliegner, J.  et al 1996, A\&A, 308, L13 
\rfr Heger, A., Langer, N., \& Woosley, S.E. 2000, ApJ, 528, 368
\rfr Hunter, I., Brott, I., Langer, N., et al. 2008a, A\&A, {\sl submitted}
\rfr Hunter, I., Brott, I., Lennon, D.J., et al. 2008b, ApJ, 676, L29
\rfr Hunter, I., Lennon, D.J., Dufton, P.L., et al. 2008c, A\&A, 479,541
\rfr Kurt, C.M. \& Dufour, R.J. 1998,RMxAC, 202
\rfr Langer, N. 1991, A\&A,252,669
\rfr Langer, N., Cantiello, M., Yoon, S.-C., et al. 2008, IAUS, 250, 167
\rfr Maeder, A., Meynet, G., Ekstrom,S., et al. 2008, {\sl these proceedings}, astro-ph/0810.0657
\rfr Mokiem, M.R., de Koter, A., Evans, C.J., et al. 2007, A\&A, 465, 1003
\rfr Morel, T., Hubrig, S., \& Briquet, M. 2008 A\&A, 481,452
\rfr Vink, J.S., de Koter, A., \& Lamers, H.J.G.L.M. 2001, A\&A, 369,574
}

\end{document}